\documentclass{PoS}
\usepackage{graphicx}
\usepackage{epsfig}
\usepackage{subfigure}
\usepackage{pstricks}

\title{The multi-faceted synergy between {\em Swift} and {\em Fermi} in radio-loud AGN studies}

\ShortTitle{The multi-faceted synergy between Swift and Fermi in radio-loud AGN studies}

\author{\speaker{F. D'Ammando}\thanks{On behalf of the Fermi Large Area Telescope Collaboration}\\
        INAF - Istituto di Radioastronomia, Via Gobetti 101, I-40129 Bologna,
        Italy \\ Dip. di Fisica e Astronomia, Universit\'a di Bologna, Viale B. Pichat 6/2, I-40127 Bologna, Italy\\
        E-mail: \email{dammando@ira.inaf.it}}
\abstract{Since its launch in 2008 June, the {\em Fermi Gamma-ray Space Telescope} has opened a new era in high-energy astrophysics. The
  unprecedented sensitivity, angular resolution and effective area of the Large Area Telescope on board {\em Fermi}, together with the nearly
  continuous observation of the entire $\gamma$-ray sky assures a formidable opportunity to study in detail $\gamma$-ray emitting AGN of various
  types. In this context the {\em Swift} satellite, thanks to its broad band coverage and scheduling flexibility, creates a perfect synergy with {\em Fermi}.

{\em Swift} and {\em Fermi} coordinated monitoring campaigns of radio-loud AGN allowed us to investigate correlated variability at different frequencies and to build time-resolved spectral energy distributions from optical to $\gamma$-rays, constraining the emission mechanisms at work in these
objects. The rapid {\em Swift} follow-up observations of $\gamma$-ray flaring AGN detected by {\em Fermi}-LAT were also fundamental in firmly associating the $\gamma$-ray sources with their low-energy counterparts. We present some interesting results obtained from {\em Fermi}-LAT and {\em Swift} observations of $\gamma$-ray flaring AGN in the first six years of {\em Fermi} operation.}

\FullConference{Swift: 10 Years of Discovery,\\
		2-5 December 2014\\
		La Sapienza University, Rome, Italy }

\begin{document}

\section{Introduction}

Only a small percentage of Active Galactic Nuclei (AGN) are radio-loud, and
this characteristic is commonly ascribed to the presence of a relativistic
jet, roughly perpendicular to the accretion disc. These jets are engines able to carry out a huge amount of power, not only in the form of radiation but especially in the form of kinetic energy and magnetic fields. Understanding the structure
and dynamics of relativistic jets is an essential step for the comprehension
and interpretation of many phenomenological properties of AGN. 
In addition to blazars and radio galaxies \cite{acero15}, recently relativistic jets
were discovered also in narrow-line Seyfert 1 galaxies (e.g.~\cite{abdo09b,
  dammando12a}). 

Multifrequency observations are a powerful tool for studying relativistic jets in radio-loud AGN.
 The {\em Fermi} satellite with its continuous
observation of the entire $\gamma$-ray sky and high sensitivity assures a
formidable opportunity to have a complete census of the extragalactic
$\gamma$-ray sky, opening a window to correlated investigations of radio-loud
AGN over the entire electromagnetic spectrum. Thanks to its broad band
coverage and scheduling flexibility, the {\em Swift} satellite creates a
perfect synergy with {\em Fermi} covering the electromagnetic spectrum from
optical to $\gamma$-rays. In this paper we present some examples of the
multi-faced synergy between {\em Swift} and  {\em Fermi} in radio-loud AGN.

\section{Long-term monitoring of AGN}

Long-term observations with {\em Swift} and {\em Fermi}, in conjunction with
other facilities in different bands are crucial for monitoring different flux levels of variable AGN and investigate if the spectral and flux variability properties are significantly different in high and
low state. The study of correlated multiwavelength variability allows us to
achieve a better understanding of the structure of the inner jet, the origin
of the seed photons for the inverse Compton process and the emission
mechanisms at work in radio-loud AGN. 

\begin{figure}
\centering
\subfigure
    {\includegraphics[width=.475\textwidth]{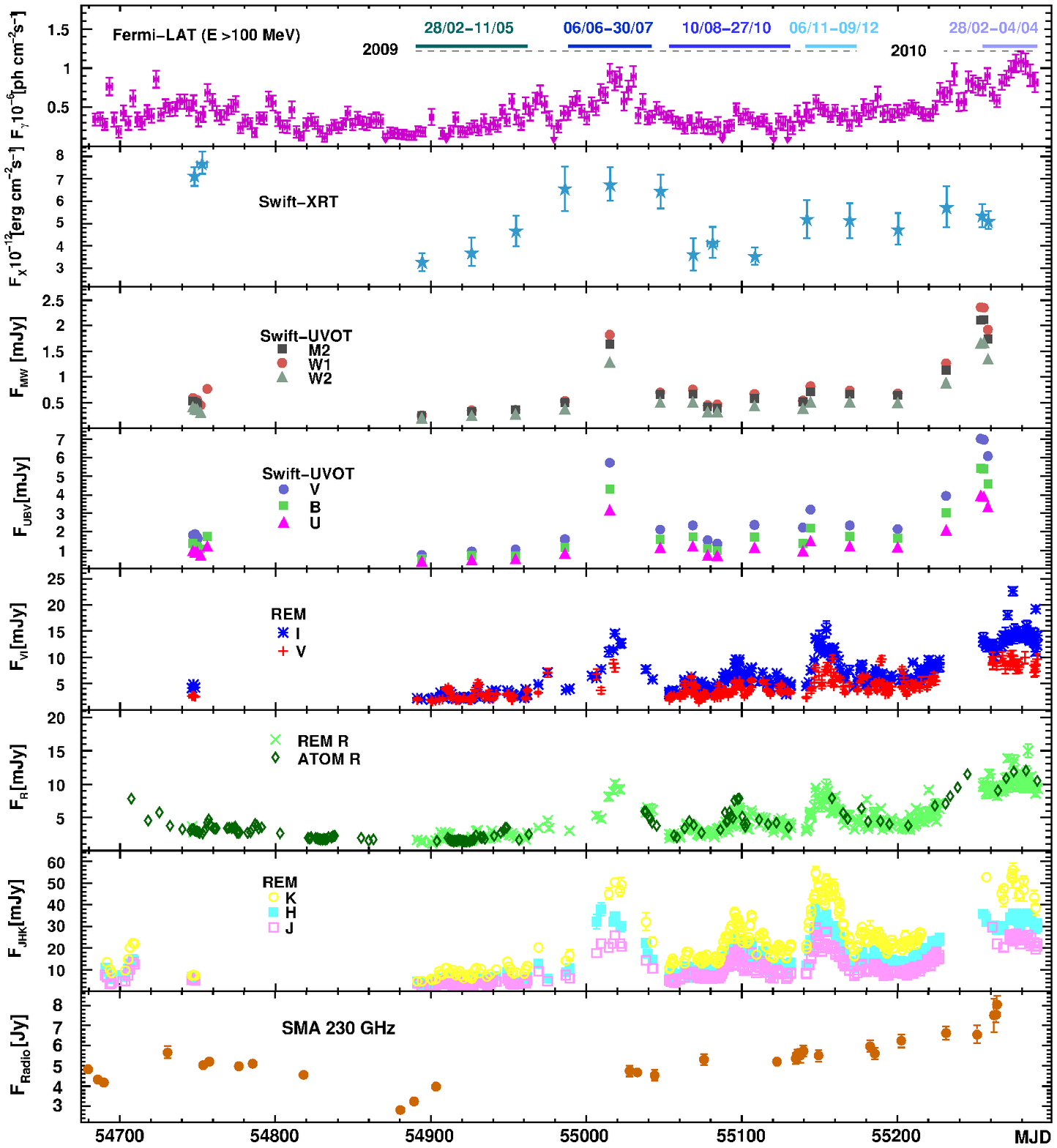}}
\hspace{4mm}
\subfigure
    {\includegraphics[width=.475\textwidth]{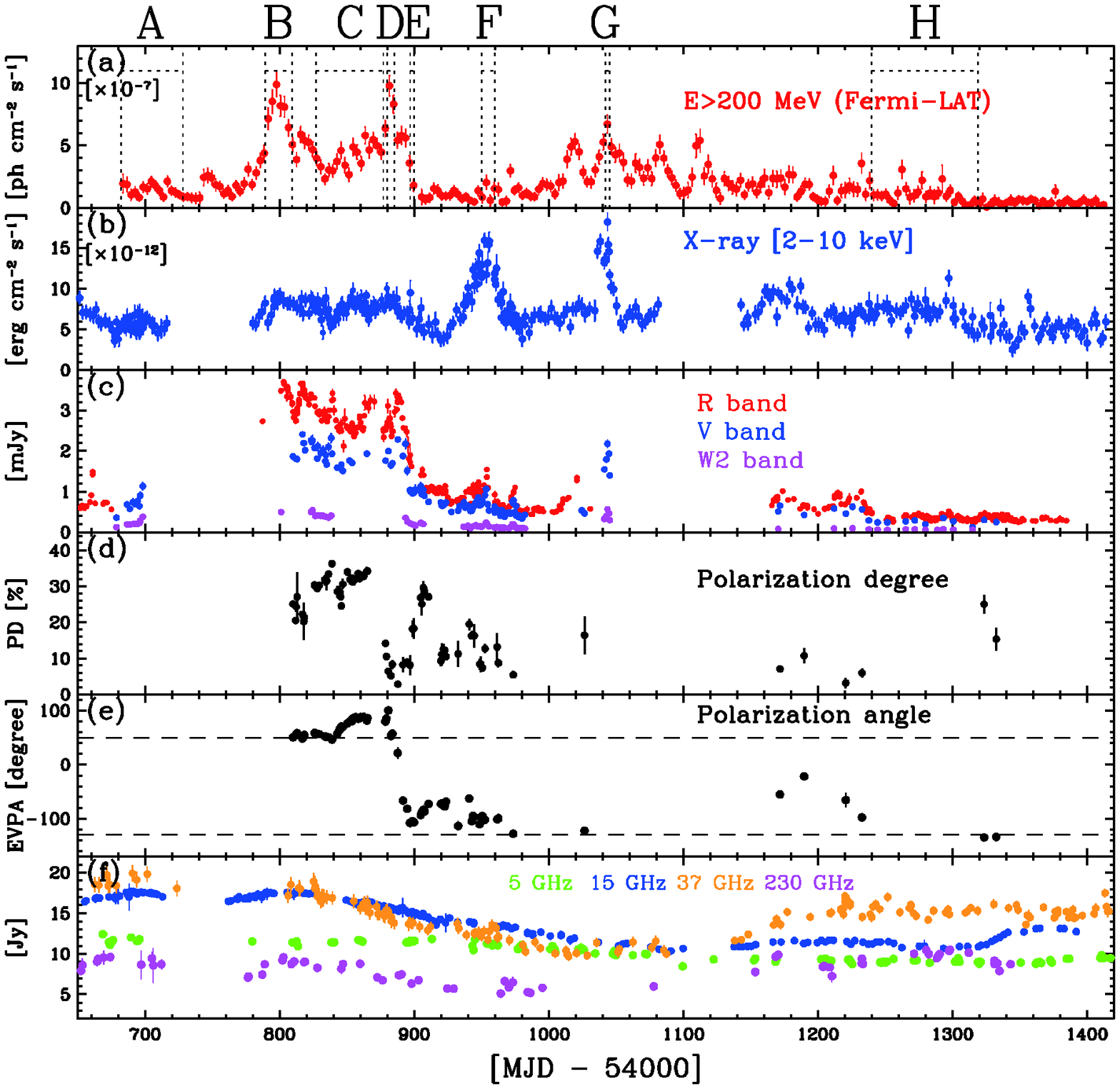}}
\caption{{\it Left Panel}: Multifrequency light curve for PKS\,0537$-$441. The period covered is
2008 August 4--2010 April 4. The
data sets were collected (from top
to bottom) by {\em Fermi}-LAT ($\gamma$-rays),
{\em Swift}-XRT (0.3--10 keV), {\em Swift}-UVOT (w1, m2, w2, and v, b, u
filters), REM (V and I bands), REM
and ATOM (R band), REM (J, H, K
bands), and SMA (230 GHz). Adapted from \cite{dammando13a}. {\it Right Panel}:
Multi-band light curves of 3C\,279 from 2008 August to 2010 August. (a):
$\gamma$-ray flux averaged over 3 days. (b): X-ray flux measured by {\em Swift}-XRT and RXTE-PCA. (c): UV-optical
fluxes in R band (red), V band (blue) and w2 band (magenta). (d): Polarization
degree in the optical band. (e): Polarization angle in the optical band. The horizontal dashed lines refer to
the angle of 50 deg and $-$130 deg. (e): Radio fluxes at 230 GHz (magenta), 37 GHz (orange), 15
GHz (blue) and 5 GHz (green). All X-ray, UV and optical data are corrected for the Galactic
absorption. Adapted from \cite{hayashida12}.}
\label{fig1}
\end{figure}

The importance of long-term monitoring was confirmed e.g. by the observations of the blazar PKS\,0537$-$441 obtained from microwaves through $\gamma$-rays by SMA,
REM, ATOM, {\em Swift} and {\em Fermi} during 2008 August--2010 April
\cite{dammando13a}. Strong variability has been observed in $\gamma$-rays,
with two major flaring episodes: 2009 July and 2010 March (Fig.~\ref{fig1},
left panel). The spectral energy distribution (SED) of the
source cannot be modelled by a simple synchrotron self-Compton model, as
opposed to many BL Lacs, but the addition of an external Compton component of
seed photons from a dust torus is needed. The 230 GHz light curve showed an
increase simultaneous with the $\gamma$-ray one, indicating co-spatiality of
the mm and $\gamma$-ray emission region likely at large distance from the central
engine. The low, average, and high activity SED of the source could be fit
changing only the electron distribution parameters.
An overall correlation between the $\gamma$-ray band with the $R$-band and
$K$-band has been observed with no significant time lag. On the other hand,
when inspecting the light curves on short time scales some differences are
evident. In particular, flaring activity has been detected in near-infrared
and optical bands with no evident $\gamma$-ray counterparts in 2009 September
and November (Fig.~\ref{fig1}, left panel). Moderate variability has been observed in X-rays with no correlation between flux and photon index. An increase of the detected X-ray flux with no counterpart at the other wavelengths has been observed in 2008 October, suggesting once more a complex correlation between the emission at different energy bands. 

\section{Isolated X-ray flares}

Similar to PKS\,0537$-$441, X-ray observations of 3C\,279 during 2008--2010 revealed a pair of pronounced
flares separated by $\sim$90 days, both not contemporaneous to increased activity in
optical or $\gamma$-ray bands (Fig.~\ref{fig1}, right panel). These isolated X-ray flares may be related to changes of
the source parameters such as the jet direction, Lorentz factor, and/or location of
the dissipation event, or they may require more unusual solutions: bulk-Compton
process, inefficient electron acceleration above a given energy, hadronic
processes \cite{hayashida12}.

\section{Identification of $\gamma$-ray sources}

\begin{figure}
\centering
\subfigure
    {\includegraphics[width=.475\textwidth]{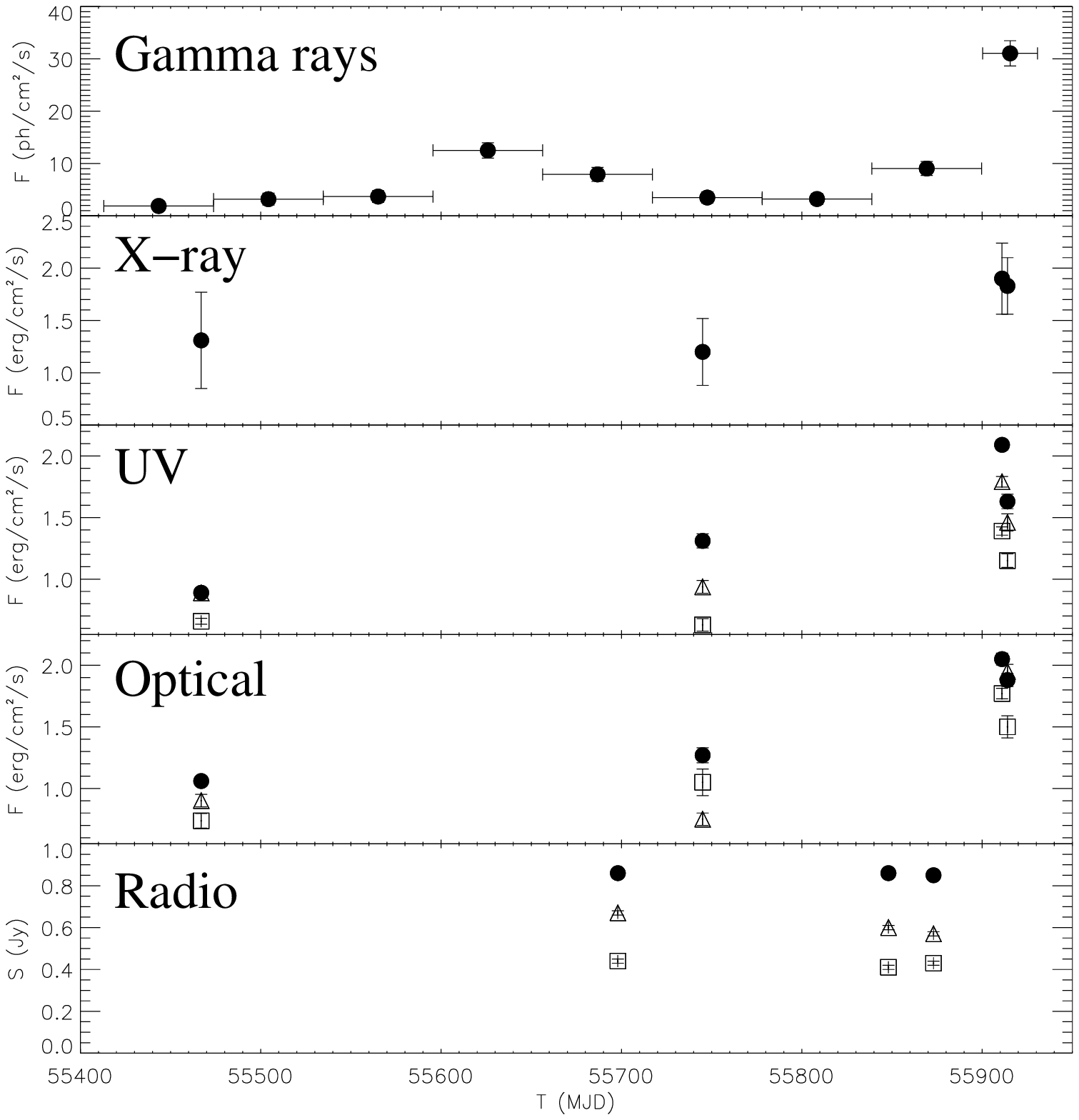}}
\hspace{4mm}
\subfigure
    {\includegraphics[width=.475\textwidth]{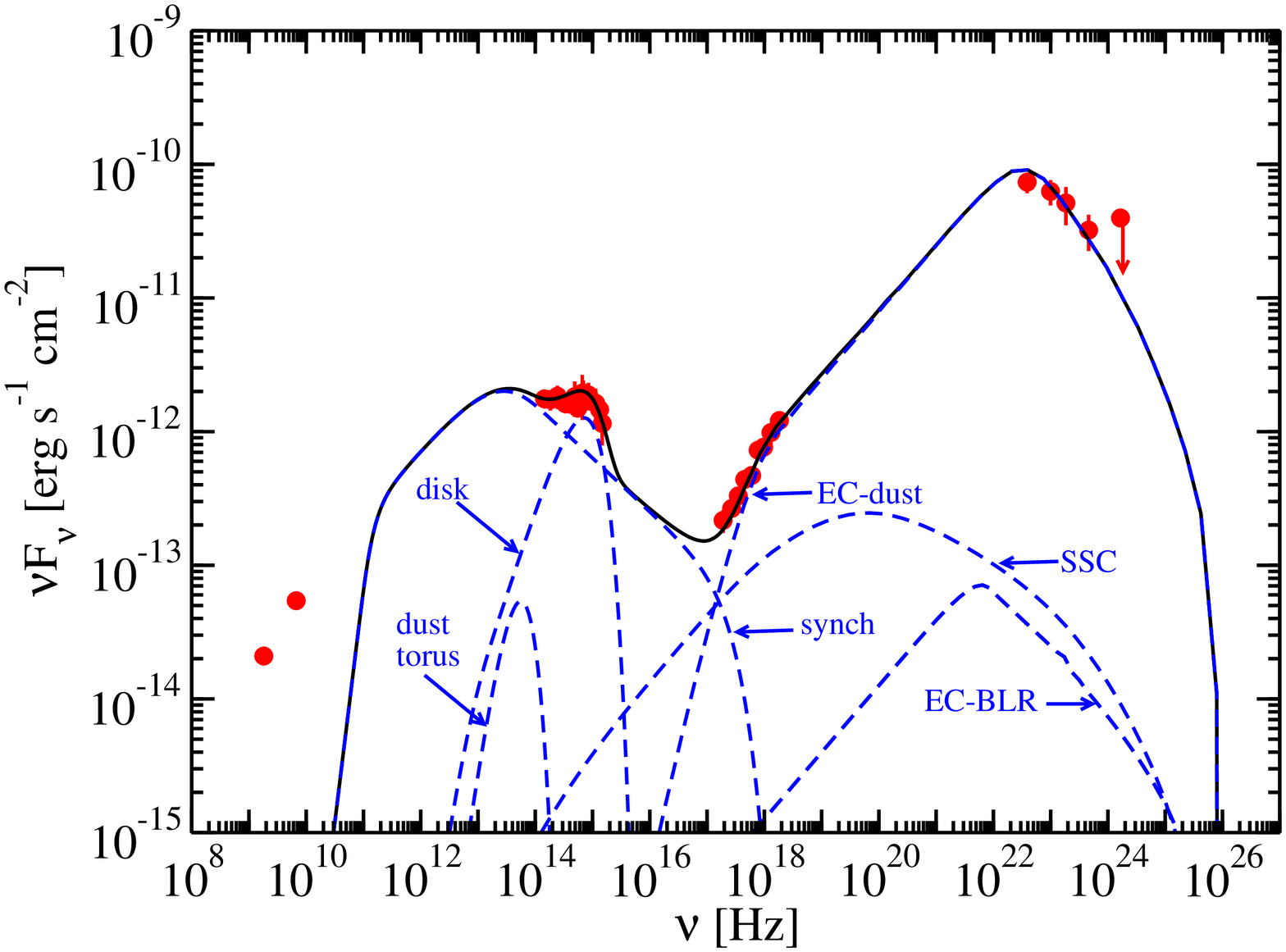}}
\caption{{\it Left panel}: light curves of PKS\,2123--463 collected in $\gamma$-rays
  by {\em Fermi}-LAT (0.1--100 GeV flux in units of 10$^{-8}$ photons cm$^{-2}$ s$^{-1}$), in
X-rays by {\em Swift}/XRT (flux in units of 10$^{-12}$ erg cm$^{-2}$ s$^{-1}$), in UV (filters w1: empty squares, m2: empty triangles and w2: filled
circles; flux in units of 10$^{-12}$ erg cm$^{-2}$ s$^{-1}$) and in optical (filters u: empty squares, b: empty triangles and v: filled circles; flux in
units of 10$^{-13}$ erg cm$^{-2}$ s$^{-1}$) by {\em Swift}/UVOT, and radio by ATCA (empty squares: 40 GHz, empty triangles: 17 GHz and filled circle:
5.5 GHz) between 2010 June and 2011 December. {\it Right panel}: SED data (circles and squares) and model
fit (solid curve) of PKS\,2123$-$463 with the model components shown as dashed curves. The data points were collected by GROND
(2011 December 18), {\em Swift} (UVOT and XRT; 2011 December 19) and {\em Fermi}-LAT (2011 December 10--19), together with radio data
from ATCA (2011 December 19) and KAT-7 (2011 December 24). Adapted from \cite{dammando12b}.}
\label{fig2}
\end{figure}

Variability is common in $\gamma$-ray blazars and provides a powerful tool to unambiguously associate them with
objects known at other wavelengths and to study the emission mechanisms at work. When combined with
simultaneous ground- and space-based multifrequency observations, the {\em Fermi}-LAT achieves its full capability
for the identification of the $\gamma$-ray sources with counterparts at lower energies and the knowledge of their
emission processes, as reported e.g. in \cite{dammando12b} for PKS\,2123$-$463 and in \cite{orienti14} for TXS 0536+135. In particular, the
strict spatial association with the lower energy counterpart together with a
simultaneous increase of the activity in optical, UV, X-ray and $\gamma$-ray
bands (Fig.~\ref{fig2}, left panel) led to a firm identification of the
$\gamma$-ray source with PKS\,2123$-$463. We fit the SED with a
synchrotron/external Compton model (Fig.~\ref{fig2}, right panel). A thermal
disc component is necessary to explain the optical/UV emission detected by {\em Swift}/UVOT. This disc has a luminosity of about
1.8$\times$10$^{46}$ erg s$^{-1}$, and a fit to the disc emission assuming a Schwarzschild (i.e., non-rotating) black hole gives a mass of about 2$\times$10$^{9}$ M$_{\odot}$. 

\section{Narrow-line Seyfert 1 galaxies}

\begin{figure}
\centering
\subfigure
    {\includegraphics[width=.475\textwidth]{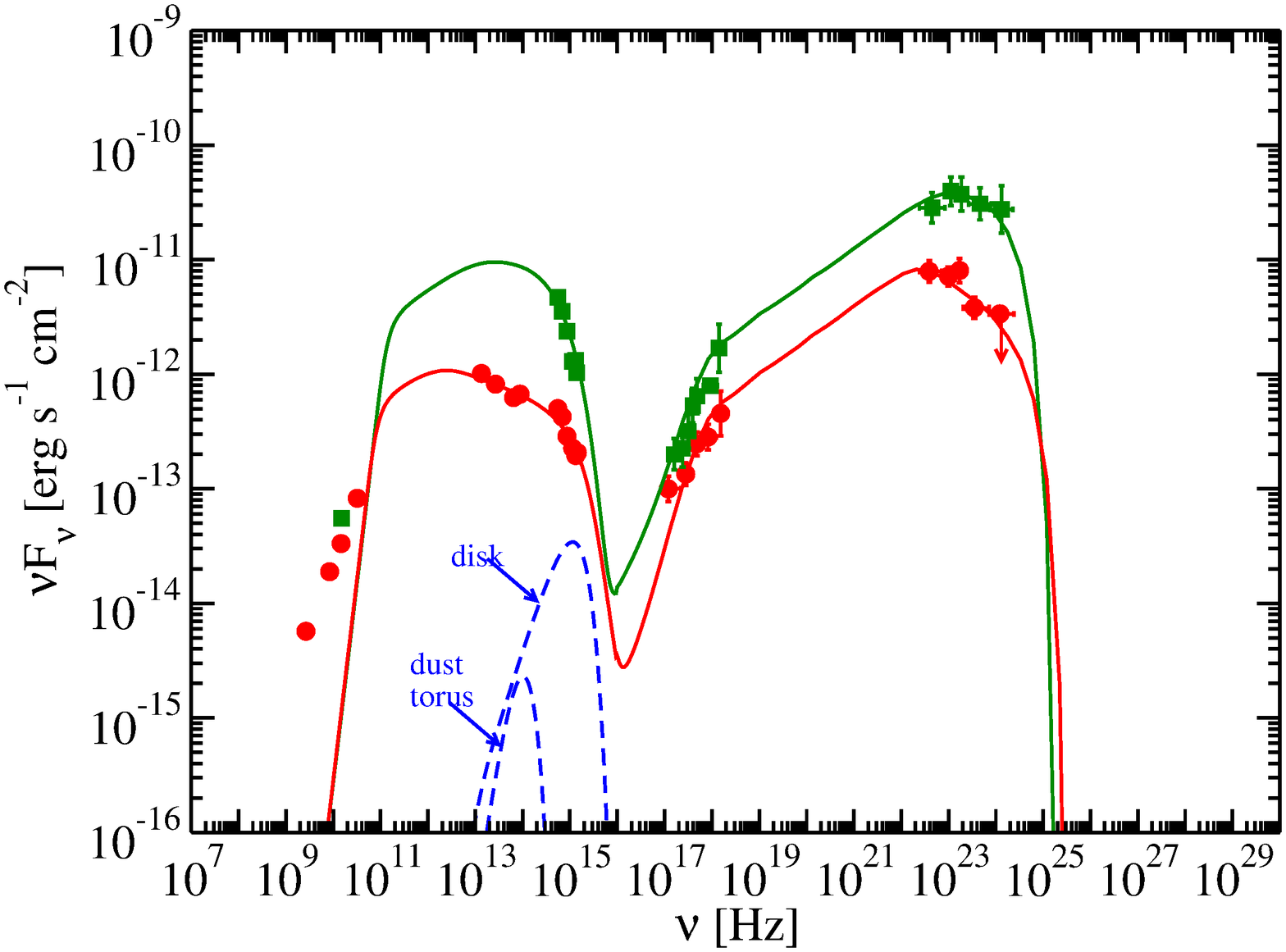}}
\hspace{4mm}
\subfigure
    {\includegraphics[width=.475\textwidth]{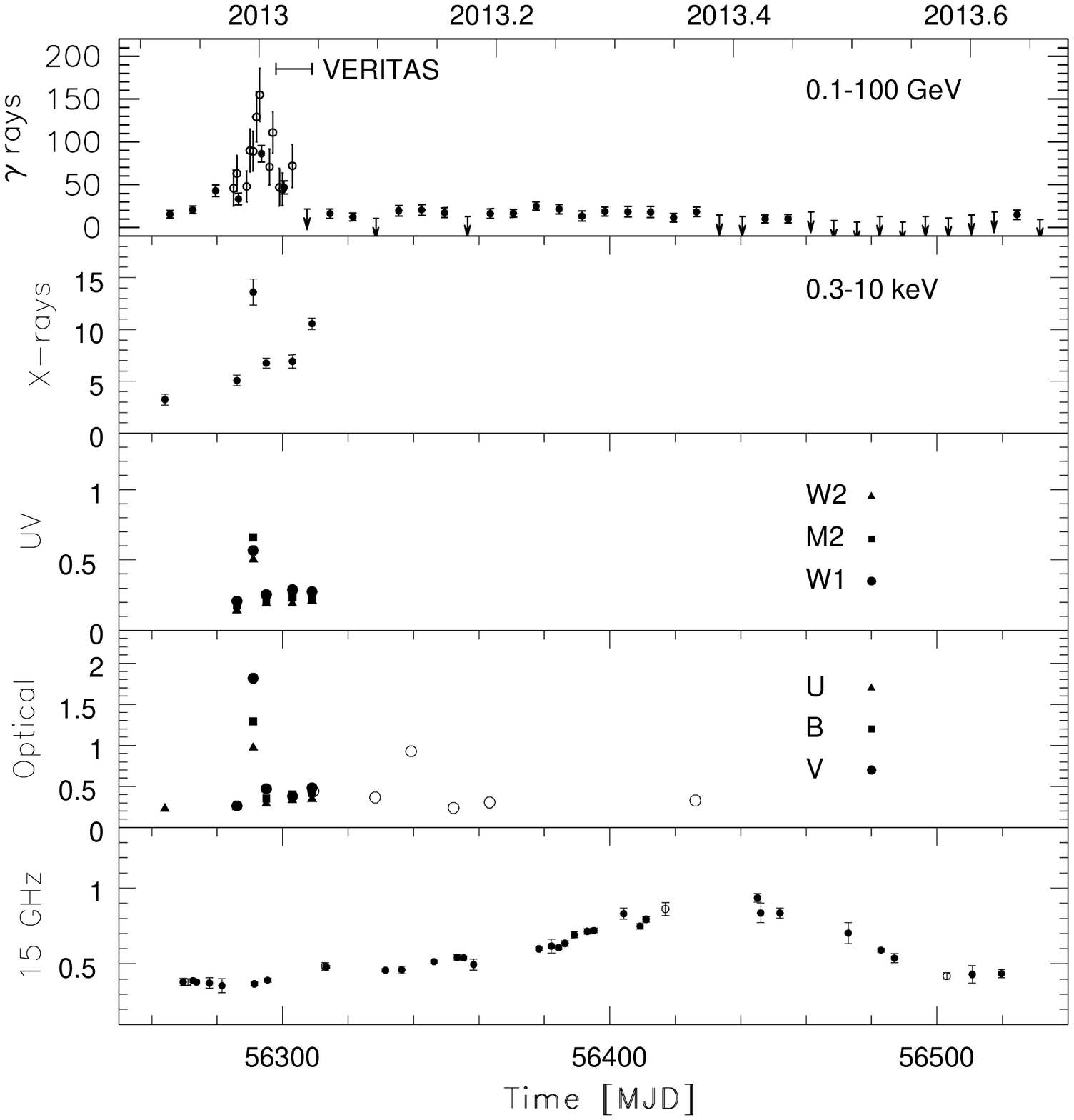}}
\caption{{\it Left panel}: SED data (squares) and
  model fit (solid curve) of SBS\,0846$+$513 in flaring activity with the
  model components shown as dashed curves. The data points were collected by
  OVRO 40-m (2012 May 17), {\em Swift} (UVOT and XRT; 2012 May 27), and {\em
    Fermi}-LAT (2012 May 20--29). The SED in the quiescent state reported in
  \cite{dammando12a} is shown as circles. Adapted from
  \cite{dammando13b}. {\it Right panel}: Multifrequency light curve for PMN\,J0948$+$0022. The period covered
  is 2012 December 1 -- 2013 August 31 (MJD 56262--56535). The data were
  collected (from top to bottom) by {\em Fermi}-LAT ($\gamma$-rays; in units
  of 10$^{-8}$ ph cm$^{-2}$ s$^{-1}$), {\em Swift}-XRT (0.3--10 keV; in units
  of 10$^{-12}$ erg cm$^{-2}$ s$^{-1}$), {\em Swift}-UVOT ($w1$, $m2$, and
  $w2$ filters; in units of mJy), CRTS (open circles; in units of mJy) and {\em Swift}-UVOT ($v$, $b$, and $u$ filters; in units of
  mJy), MOJAVE (open circles) and OVRO (15 GHz; in units of Jy). The
  horizontal line in the top panel indicates the period of the VERITAS
  observation. Adapted from \cite{dammando15}.}
\label{fig3}
\end{figure}

Thanks to {\em Swift} and {\em Fermi}-LAT observations, we are able to
investigate the broad-band spectra and the correlated variability in different
energy bands of this new class of $\gamma$-ray emitting AGN \cite{abdo09a, abdo09b, dammando12a}.
We compared the SED of SBS\,0846$+$513 during the flaring state in 2012 May with that of a quiescent
state (Fig.~\ref{fig3}, left panel). The SED of the two different activity states, modeled by an external Compton component of seed
photons from a dust torus, could be fitted by changing the electron distribution parameters as well as the
magnetic field \cite{dammando13b}, consistent with the modeling of different activity
states of PKS\,0208-512 \cite{chatterjee13}. A significant shift of the synchrotron peak to
higher frequencies was observed during the 2012 May flaring episode, similar
to FSRQ (e.g., PKS\,1510-089; \cite{dammando11}). Contrary to what is observed
in PMN J0948$+$0022 \cite{dammando14}, no significant
evidence of thermal emission from the accretion disc has been observed in SBS\,0846$+$513.
A complex connection between the radio and $\gamma$-ray emission was observed
for SBS\,0846$+$513 and PMN\,J0948$+$0022 (Fig.~\ref{fig3}, right panel), as discussed in detail e.g. in
\cite{dammando13b, orienti13, dammando14, foschini12}. At Very High Energy
(VHE; E $>$ 100 GeV), VERITAS observations of PMN\,J0948$+$0022 were carried out during 2013 January 6--13, after the $\gamma$-ray flare observed by {\em Fermi}-LAT on 2013 January 1. These observations resulted in an upper limit of F$_{> 0.2\rm\,TeV}$ $<$ 4$\times$10$^{-12}$ ph cm$^{-2}$ s$^{-1}$ \cite{dammando15}.

\section{Constraining the BH spin of FSRQ}

The fit of the optical/UV part of the spectrum of the flat spectrum radio
quasar 4C\,$+$21.35 collected by {\em Swift}/UVOT on 2010 April 29 (mainly produced by the accretion disc) seems to
favor an inner disc radius of $<$6 gravitational radii, as one would expect from a
maximally prograde-rotating Kerr black hole \cite{ackermann14}.

\begin{figure}
\centering
\includegraphics[width=.5\textwidth]{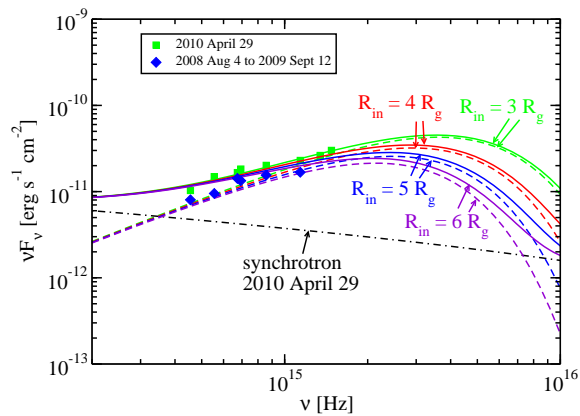}
\caption{ Model disc emission for several inner disc radii are shown (dashed curves), while the synchrotron
component from the model fit of 2010 April 29 is shown as the dot-dashed curve. The total
(synchrotron + disk) emission is shown as the solid curves. Models for 4C\,$+$21.35 with large inner disc radii do not
provide an adequate fit to the UV data of 2010 April 29 \cite{ackermann14}.}
\label{fig4}
\end{figure}

\begin{acknowledgments}
The \textit{Fermi}-LAT Collaboration acknowledges support for LAT development, operation and data analysis from NASA and DOE (United States), CEA/Irfu and IN2P3/CNRS (France), ASI and INFN (Italy), MEXT, KEK, and JAXA (Japan), and the K.A.~Wallenberg Foundation, the Swedish Research Council and the National Space Board (Sweden). Science analysis support in the operations phase from INAF (Italy) and CNES (France) is also gratefully acknowledged.
\end{acknowledgments}

\end{document}